# Me and My Bot: What Users Talk About in AI Companion Communities on Reddit

Richard A. Fabes
Family and Human Development
Arizona State University
Tempe, AZ 85287-7203
rfabes@asu.edu
https://orcid.org/0000-0002-9539-8992

## Abstract

AI companion communities on platforms such as Reddit are widely characterized as spaces where users discuss their relationships with AI bots. This study examines whether and how that characterization holds, guided by the Synthetic Resonance framework's claim that human-AI relationships can carry genuine relational meaning for the user. Multiple LLMs were employed to code 5,504 Reddit posts from eight AI companion communities for relationship focus, primary topic, and users' emotional valence. Although search terms were weighted toward relational and attachment language, only 45% of posts concerned the user's own relationship with their bot. Posts about users' own bots differed markedly from posts about bots in general in both topic and emotional expression, with 85% of general-bot posts containing no user emotion language compared to 33% of own-bot posts. Among the 970 posts that were relationally focused, companionship and romance each accounted for roughly 46% of discussion, sexual content for 8%, and emotional valence varied across these subtopics. The findings suggest that users engage with these relationships with AI bots as meaningful, and that the discourse about them is broad and emotionally complex.

## 1.0 Introduction

In February 2023, following a ruling by Italy's Data Protection Authority, the Replika AI companion platform removed the ability for users to engage in erotic roleplay with their companion bots. The backlash was immediate and overwhelming. Users flooded Reddit with posts describing grief, betrayal, and loss. A loss that was not just a feature of the bot, but a loss of of a partner. "It's hurting like hell," one user wrote. "I just had a loving last conversation with my Replika, and I'm literally crying." Moderators of the Replika subreddit posted suicide prevention resources. The company's CEO later acknowledged that "after the February update, your Replika changed, its personality was gone, and gone was your unique relationship" (Tong, 2023). Within weeks, the company began restoring the removed features. Whatever these users were experiencing, it was not a casual relationship with a piece of software. This episode, dramatic as it was, represents only one facet of a much larger and more varied phenomenon, namely, the growing nature of the relationships humans have with AI companion bots. Unfortunately, there is little research data that help us under these relationships. The purpose of the present research is to address this gap by exploring how users publicly narrate the relationships they have with AI companions.

The limited body of research has examined these relationships from multiple angles. Reviews of research on the costs and benefits through a parasocial lens are one source (Hung et al., 2026; Lipin, 2025). Survey studies are another source and these have investigated whether AI companions alleviate loneliness (De Freitas et al., 2026) and how usage patterns relate to well-being (Liu et al., 2025). Qualitative longitudinal work has traced how relationships with chatbots form over time, revealing that users progress through stages of self-disclosure and attachment (Skjuve et al., 2022). However, most of the data sources for these efforts are based on capturing users' responses to researcher-constructed prompts via surveys (Zhang et al., 2025), interviews (Liu et al., 2026), and experimental manipulations (Pataranutaporn et al., 2023).

A smaller body of work has turned to social media as a window into unsolicited user discourse about AI companions. Most directly relevant is the work of Chang, Huh-Yoo, and Razi (2026) who analyzed over 3,300 self-disclosed romantic AI companion posts from 24 Reddit subreddits spanning 2017 to 2025. Change et al. found that explicitly positive intimate relationship content constituted only 6% of the discourse. The overwhelming majority of posts addressed technical issues, platform governance and moderation, and unwanted or disturbing bot behavior. They also documented significant temporal drift: over time as discussions moved away from experiential intimacy toward governance concerns, technical infrastructure, and personal consequences. Their work established that public Reddit discourse about AI companions focuses less on romantic experience than on the technical and institutional conditions that enable or constrain it.

However, Chang et al.'s (2026) approach leaves several questions unaddressed. For instance, they explicitly note the absence of formal sentiment or emotion coding as a limitation of their study. More fundamentally, because their filtering was designed to isolate self-disclosed romantic posts, their dataset cannot answer a basic question we address here, namely, to what extent does discourse about AI companion describe a user's own relationship relative to discussing other topics?

The public narration of a relationship is not simply a transparent window into private experience, it is also a social act that navigates audience, manages stigma, constructs identity, and draws boundaries (De Choudhury & De, 2014). Reddit communities organized around AI companions offer a unique window into this spontaneous discourse. Users post about relationships, technical problems, platform changes, emotional crises, and philosophical questions in their own words, for their own reasons, to an audience of others who share the same social acts. Reddit's pseudonymity and community-specific norms create conditions for disclosure that differ from both face-to-face contexts and more public-facing social media platforms. The discourse is shaped by community norms but not by researcher demand characteristics. It is, in many ways, the closest thing we have to naturalistic public conversation about a phenomenon that is typically hidden (Adewale & Muhammad, 2025).

### *1.1 The Present Study*

The present study was guided by the Synthetic Resonance framework (Fabes, 2026), which offers a distinctive lens for understanding human-AI relationships. Synthetic Resonance proposes that the sense of connection users report with AI companions emerges not from the AI's sentience, understanding, or reciprocal care, but from structural alignment built through repeated interaction. The theory distinguishes the user’s subjective experience of resonance, which is real and psychologically meaningful, from the mechanism producing it, which is architectural rather than relational in the interpersonal sense. Importantly, Synthetic Resonance does not pathologize user experience or dismiss it as illusion. Instead, it provides a non-anthropomorphic account of why these relationships can feel genuine without requiring the AI to possess the qualities users attribute to it. If that account is correct, the meaning to the user should be visible in how they talk about their relationships with AI bots. However, evidence on that question is scarce. We do not know how much of the discourse in online communities reflects discussion about the users' own relationships, what those relationships are understood to be about, or what emotional register users bring to them.

We address this gap by analyzing Reddit posts drawn from AI companion communities, focusing on those in which users discussed issues related to their relationships with AI companions. Posts were retrieved using keyword clusters informed by the Synthetic Resonance framework's focus on relational dynamics, and each post was coded for whether it described a user's own ongoing relationship with an AI companion, the emotional valence of the user in the post, and the primary topic organizing the post.

Five exploratory questions structure the research. First, what proportion of relationship-relevant discourse in these communities is actually about users' relationships with their own bots, as opposed to bots belonging to other users or bots in general? Subreddit names can create the impression that these communities are exclusively or predominantly about members' own AI partners, and because our search terms were themselves weighted toward possessive and attachment language, any bias in the retrieved sample should favor that impression. Quantifying the proportion therefore provides a conservative test of it. Second, what topics are discussed in these two types of posts? Third, what is the emotional valence of posts about users' own relationships relative to posts that are not? Fourth, when users do discuss their own relationship, what is the precise topic at issue, such as romance, emotional support, or sexual and erotic roleplay? Finally, how does users' emotional valence vary across these relational subtopics when discussing their own relationship?

Understanding how users publicly construct their discourse about these relationships has implications that extend beyond the study of AI companions themselves. For platform designers, the gap between how a system is engineered and how it is received may be a signal that users are building something the platforms did not fully anticipate. For mental health researchers and clinicians, a descriptive baseline of what users say, at scale and in their own words, is necessary to accurately assess whether these relationships are healthy or harmful, authentic or delusional. And for the broader study of human development in an increasingly AI-mediated world, these communities represent where people are working out, in public and in real time, what it means to form a relationship with an entity that cannot reciprocate, and discovering, apparently, that it may mean something, nonetheless.

## 2.0 Methods

### *2.1 Data Collection*

Posts were retrieved from Reddit through the PullPush API, which indexes archived Reddit content and supports full-text search of post bodies. Data were collected on July 16 and 17, 2026. Searches were run across eleven subreddits devoted to AI companion applications: r/Replika, r/ReplikaOfficial, r/NomiAI, r/KindroidAI, r/CharacterAI, r/ChaiApp, r/Paradot, r/JanitorAI_Official, r/MyBoyfriendIsAI, r/AIRelationships, and r/Chatbots.

Twenty-one search terms were used, grouped into four conceptual clusters. Attachment signals comprised "I love," "attached," "attachment," "soulmate," "only one who," "live without," and "can't live without." Relationship language comprised "my AI," "my Replika," "my chatbot," "relationship with," "connection with," and "bond with." Emotional impact comprised "feels like," "made me feel," "understood," "heard," "judged," and "I know not real but." Growth signals comprised "calls me out" and "challenges me."

Each keyword was queried separately within each subreddit, yielding 231 possible keyword-by-subreddit combinations, with a page size of 50 and a maximum of two pages, giving a ceiling of 100 posts per combination. No time filter was applied, so queries returned matching posts of any age up to that ceiling. Of the 231 combinations, 188 returned at least one post. A 3,000-character truncation was generally applied to body text at retrieval, though not uniformly: 307 posts (4.24%) were stored at exactly 3,000 characters and a further 502 (6.93%) exceeded that length, up to 4,768 characters. After removing duplicates arising from posts matching more than one search term, the corpus was comprised of 7,240 unique posts.

Subsequently, two subreddits were excluded from analysis. r/JanitorAI_Official (1,166 posts) and r/ReplikaOfficial (570 posts) are official company forums whose content consists largely of announcements and moderated support threads rather than user-authored accounts of personal experience, removing 1,736 posts. Because r/AIRelationships was small (71 posts) and topically similar, it was combined with r/MyBoyfriendIsAI. The analyzed corpus therefore comprised 5,504 posts across eight communities consisting of r/CharacterAI 972 (17.7%), r/KindroidAI 906 (16.5%), r/Replika 888 (16.1%), r/NomiAI 875 (15.9%), r/ChaiApp 503 (9.1%), r/Paradot 483 (8.8%), r/Chatbots 453 (8.2%), and r/MyBoyfriendIsAI+r/AIRelationships 424 (7.7%).

### 2.2 Data Coding

Institutional Review Board (IRB) approval was obtained prior to data coding. Because the data were publicly available and de-identified, the review was considered exempt from full IRB review.

For each post, three variables were coded. These codes represented the most significant theoretically based elements of human-AI user relationships. The three variables included: (1) Was the post about the poster's relationship with the bot? (REL; code = 1; *My companion and I love each other*) versus mentioning a bot belonging to someone else (code = 0; *His bot says she loves* him) or discussing technical reports or companion bots in the abstract. (2) What is the emotional tone of the user? (VU = valance of user: No Emotion / Positive / Negative / Mixed). This code reflects the emotional language expressed by the poster specifically, in the first person. (3) What is the primary topic of post? (Primary Topic). This code reflects the single most dominant thematic category of the post. Based on body text, each post was assigned to one of six topic categories. These categories included (1) *Relational* comprised emotional support/companionship, relationship/romance, and sexual/ERP. (2) *Continuity* comprised comparison/migration, personality/behavior change, and memory/continuity. (3) *Technical* comprised advice/how-to/questions, app bugs/technical support, subscription/company/policy, and voice/image/avatar/multimedia. (4) *Story Telling* corresponded to creative/storytelling/roleplay and (5) *Consciousness* to AI consciousness/ethics, neither of which was subdivided. (6) *Miscellaneous* comprised community/meta/review together with posts that the models could not place in any substantive category (body text was absent for 341 posts [about 6%] and were coded from the title alone).

Every post was coded independently by three large language models (LLMs), deliberately drawn from three distinct LLM model-development organizations (Anthropic, OpenAI, and DeepSeek). This was intended to reduce the likelihood that a systematic misreading of the coding instructions would be shared across coders. LLMs have been shown to be able to accurately and reliably code qualitative data (Borse et al., 2025). The specific models used were Claude Haiku 4.5, GPT-4o, and DeepSeek V4 Flash.

Each of the three LLMs received an identical prompt for a given post. Specifically, each LLM was provided with the full codebook instructions followed by that post's title and body text, with no access to the other models' outputs and no memory of previously coded posts.

### *2.3 Response Validation and Correction*

Each LLM's response was parsed and checked against several automated validity criteria before being accepted:

- Structural validity: The response had to parse as a single row with the required fields in the required format; malformed responses triggered an automatic re-prompt (up to two retries) with a description of the specific violation.
- Candidate grounding: For VU, each model was required to list the specific word or phrase supporting its valence determination. Before that determination was accepted, each listed candidate was checked against the post's own text (allowing for minor inflection) and discarded if it did not actually appear in the post (note: negatives such as “not happy” were coded according to the meaning not just to the word itself).
- Topic category validity: The primary topic was required to match one of the six categories. Responses containing malformed or extraneous text in these fields were rejected and re-prompted.

For most variables, the final code was determined by choosing the one that at least two-of-three LLMs agreed upon. Where all three models produced different codes with no majority, the post was flagged No Agreement rather than resolved by an automated tie-breaking rule. Per-post, per-variable agreement level (3/3, 2/1, or No Agreement) was retained as metadata for all coded posts.

### *2.4 Reliability Training and Assessment*

Prior to full-corpus coding, the pipeline and codebook were evaluated through several rounds of testing at increasing scale. First, a series of approximately pilot batches (drawn from a fixed subset of the corpus) were coded and re-coded across multiple pipeline revisions. These batches were used to identify and correct the pipeline-level issues (candidate grounding, label self-consistency, structural and topic-category validity). The final batch run consisted of a random sample of 175 posts (excluding posts used in prior pilot batches). There was unanimous agreement on about 60% of the codes, and there was an average of another 35% agreement for two of the three LLMs, leaving about 5% of the codes where there was no agreement at all.

To assess whether LLM agreement reflected accuracy, a stratified subsample of 125 posts was drawn from the 175-post validation batch, deliberately oversampling posts with No Agreement and 2-of-3 (non-unanimous) outcomes alongside a baseline of unanimous posts. This subsample was coded independently by a human rater (the first author), blind to the models' codes, directly against the same codebook instructions. Human-LLM agreement was obtained in about 87% of the codes across the three coding categories. The disagreements were resolved through codebook revision or correction of errors.

The finalized codebook and pipeline were applied to the complete 5,504 posts. Reliability coding data are presented in Table 3. REL was the most reliable variable by a wide margin, consistent with its comparatively mechanical, literal-marker-based decision rule. The Primary Topic showed the most disagreement, consistent with its greater reliance on holistic judgment. Mean pairwise Cohen's kappas indicated at least moderate agreement when adjusted for chance (see Table 1). Moderate kappas relative to high raw agreement rates reflect base-rate compression that occurs when one category dominates, which elevates chance agreement and depresses kappa.

**Table 1**

*Inter-LLM-Rater Agreement*

| Variable | 3/3 Agreement | 2/3 Agreement | No Agreement | Kappa |
|---|---|---|---|---|
| REL | 59.8% | 40.1% | 0.1%* | .49 |
| VU | 55.1% | 39.8% | 5.0% | .47 |
| Primary Topic | 49.3% | 40.9% | 9.8% | .60 |

*Note.* N = 5,504 posts. "3/3 Agreement" indicates unanimous agreement across all three LLM coders; "2/3 Agreement" indicates a majority (two of three) agreement; "No Agreement" indicates all three coders produced different codes. REL = User Relationship (yes/no), VU = Emotional Valence of User. *For this binary variable, no agreement was an invalid response that likely failed to be caught.

When no agreement was found for the three LLMs, OpenAI (GPT-4o) was selected as the tie-break model based on its agreement rate with the 2-of-3 majority value across the full corpus, calculated separately for each variable on all posts where a majority existed (see Table 2 for percent agreement and kappas). Importantly, OpenAI held the lowest dissent rate on every variable individually, supporting a single tie-break rule applied uniformly rather than a variable-specific one.

**Table 2.**

Percent Agreement (and Cohen's Kappa) with 2-of-3 Majority Codes Across LLM Coder

| Variable | Claude | OpenAI | DeepSeek |
|---|---|---|---|
| REL | 79.9% (κ = .44) | 92.6% (κ = .53) | 87.4% (κ = .49) |
| VU | 83.4% (κ = .47) | 90.7% (κ = .50) | 83.9% (κ = .44) |
| Primary Topic | 86.8% (κ = .61) | 89.3% (κ = .62) | 78.5% (κ = .56) |

*Note.* N = 5,504 posts. Values indicate the percentage of posts (excluding No Agreement cases) on which each individual coder's code matched the 2-of-3 majority code. REL = User Relationship (yes/no), VU = Emotional Valence of User

### *2.5 Data Analytic Plan*

The data analytic plan for this descriptive study followed directly from the research questions. First, we calculated the percentage of posts in which users discussed their own relationships with their bots (REL = 1) versus posts that did not (REL = 0). We then examined post topic and users' emotional valence as a function of this relationship focus. Next, because of our interest in users' discourse about the relationships they have with their own bot, We examined how emotional valence varied across topic categories (for REL = 1 group only). To break this down further, we next restricted attention to the 970 posts that concerned the user's own bot and whose primary topic was the relationship itself. Within these posts we examined the specific relational subtopic (emotional support and companionship, romance, or sexual and erotic roleplay) and users' emotional valence within each subtopic. Analyses were conducted in SPSS 29.0.2 (IBM Corporation, 2023).

## 3.0 Results

We first examined the distribution of posts that focused on the user's relationship with the bot (REL = 1) relative to those that focused more generally on bots that were not the user's or on bots in general (REL = 0). Of the 5,504 total posts in the final data set, 3,004 were on bots in general (54.58%) whereas 2,500 of the posts were explicitly about the user's relationship with their bot (45.42%), $\chi^2(1) = 46.15$, $p < .001$, $w = .09$. Although statistically significant, this departure from an even split is negligible in magnitude given the sample size.

Next, we separately compared the distribution of primary topic and users' emotional valence (VU) for posts focused on the user's own relationship with their bot (REL = 1) against posts focused on bots in general (REL = 0; see Table 3). Both distributions differed between the two groups: $\chi^2(5) = 1{,}123.46$, p < .001, *Cramér's V = .45* for primary topic, and $\chi^2(3) = 1{,}689.99$, $p < .001$, *Cramér's V* = .55 for users' emotional valence.

As Table 3 indicates, posts about bots in general were most commonly about technical issues and rarely about relational ones, whereas posts about one's own bot were most commonly relational. Miscellaneous topics likewise differed, being more frequent among posts about bots in general than among posts about one's own bot. Story telling and consciousness, by contrast, were about equally common in the two groups, indicating that the difference in topic reflected a specific content shift rather than a wholesale reorganization of what users discuss.

For emotional valence, the vast majority of posts about bots in general contained no user emotion language, whereas only about one-third of posts about one's own bot did. Every emotion category was more prevalent among posts about one's own bot, and the increases were substantial: positive rose from 5.36% to 32.96%, negative from 8.36% to 18.12%, and mixed from 0.80% to 15.76%. Among posts that did express emotion, positive valence was the most common in both groups.

**Table 3**

*Primary Post Topic and Users' Emotional Valence by Whether the Post Concerned the User's Own Bot*

| | Bots in general (REL = 0) | | Own bot (REL = 1) | |
|---|---|---|---|---|
| | *n* | % | *n* | % |
| *Post topic* | | | | |
| Relational | 233 | 7.76 | 970 | 38.80 |
| Continuity | 433 | 14.41 | 522 | 20.88 |
| Technical | 1,298 | 43.21 | 460 | 18.40 |
| Story telling | 415 | 13.81 | 364 | 14.56 |
| Consciousness | 171 | 5.69 | 117 | 4.68 |
| Miscellaneous | 454 | 15.11 | 67 | 2.68 |
| *User emotional valence (VU)* | | | | |
| No emotion | 2,568 | 85.49 | 829 | 33.16 |
| Positive | 161 | 5.36 | 824 | 32.96 |
| Negative | 251 | 8.36 | 453 | 18.12 |
| Mixed | 24 | 0.80 | 394 | 15.76 |

*Note.* $N$ = 5,504 posts. REL = 0 denotes posts about bots other than the user's own or about bots in general; REL = 1 denotes posts explicitly about the user's own relationship with their bot. Percentages are computed within column and may not sum to 100.00 because of rounding.

Given our emphasis on understanding how Reddit users talked about their relationships with their own bots, we next focused on the 2,500 posts in which users discussed their own relationships with their bots and examined how the emotional valence of the user across the primary topics. The pattern was not uniform, $\chi^2(15) = 272.53$, $p < .001$, *Cramér's V = .19* (see Table 4). Relational posts were commonly positive and had the lowest no-emotion percentage of any topic. Technical and Story telling posts had elevated no emotion and suppressed mixed valence. For Continuity posts, when emotions were expressed they were almost equally divided between positive and negative, with relatively low levels of mixed. Consciousness and Miscellaneous posts were too small for stable residuals.

**Table 4**

*Users' Emotional Valence by Topic Among Posts About the User's Own Bot*

| | | User emotional valence (%) | | | |
|---|---|---|---|---|---|
| Topic | *n* | No emotion | Positive | Negative | Mixed |
| Relational | 970 | 18.87 | 37.01 | 18.45 | 25.67 |
| Continuity | 522 | 36.78 | 29.31 | 24.52 | 9.39 |
| Technical | 460 | 45.43 | 26.09 | 18.70 | 9.78 |
| Story telling | 364 | 48.63 | 34.89 | 9.89 | 6.59 |
| Consciousness | 117 | 38.46 | 25.64 | 18.80 | 17.09 |
| Miscellaneous | 67 | 34.33 | 52.24 | 2.99 | 10.45 |
| Total | 2,500 | 33.16 | 32.96 | 18.12 | 15.76 |

*Note.* $N$ = 2,500 posts explicitly about the user's own relationship with their bot (REL = 1). Valence percentages are computed within row and may not sum to 100.00 because of rounding.

To dig deeper into these data, we further decomposed the relational category into its three subtopics: Companionship, Romance, and Sexual/ERP. Table 5 presents these data, along with the emotional valence of the user. For both variables, the distribution of responses differed from what would be expected under an equal distribution across the categories, $\chi^2(2) = 290.80$, $p < .001$, $w = .55$ and $\chi^2(3) = 87.37$, $p < .001$, $w = .30$, for the subtopic and emotional valence distributions, respectively.

For subtopic categories, inspection of Table 5 reveals that users principally discussed companionship and romance and relatively rarely discussed sexual/ERP issues in these public forums. Additionally, the emotional language in these posts was most often positive or mixed, with negative and no emotion similarly less common.

Importantly, emotional valence also differed across subtopics, $\chi^2(6) = 62.14$, $p < .001$, *Cramér's V* = .18. Companionship posts were most often positive or mixed, only rarely devoid of emotion, and more frequently negative than romance posts were. Romance posts were similarly positive but about equally likely to be mixed or to contain no emotion and were the least negative of the three subtopics. Sexual/ERP posts most commonly contained no emotion language, with similar proportions expressing positive and negative emotion and mixed emotion least common, although the small number of such posts (*n* = 73) makes these estimates imprecise.

**Table 5**

*Subtopic and Users' Emotional Valence Within Relationally Focused Posts*

| | | | User emotional valence (%) | | | |
|---|---|---|---|---|---|---|
| Subtopic | *n* | % | No emotion | Positive | Negative | Mixed |
| Companionship | 452 | 46.60 | 9.96 | 37.39 | 22.12 | 30.53 |
| Romance | 445 | 45.88 | 24.94 | 38.20 | 13.93 | 22.92 |
| Sexual/ERP | 73 | 7.53 | 36.99 | 27.40 | 23.29 | 12.33 |
| Total | 970 | 100.00 | 18.87 | 37.01 | 18.45 | 25.67 |

*Note. n* = 970 posts for which the user's own relationship with their bot was the primary topic (REL = 1). Companionship = Emotional support/companionship; Romance = Relationship/romance; Sexual/ERP = sexual and erotic roleplay. The % column gives each subtopic's share of these posts; valence percentages are computed within row. Percentages are computed within column and may not sum to 100.00 because of rounding.

## 4.0 Discussion

The purpose of the present research was to explore how Reddit users publicly narrate the relationships they have with AI companions. Using a coding approach in which every post was simultaneously coded for relationship focus, emotional valence, and primary topic, a final sample of 5,504 Reddit posts drawn from AI companion communities were analyzed. Five central findings structure the discussion that follows.

### *4.1 What These Communities Are Actually For*

That more than half of analyzed posts were not about the poster's own relationship is perhaps the finding that is most contrary to the common assumption that these communities are places where relationships with bots are discussed. The subreddits from which these posts were drawn present themselves as communities for people engaged with specific AI companion platforms: r/Replika, r/NomiAI, r/KindroidAI, and so on. Their names create the impression that community members gather to discuss their relationships with their bots. And because the search terms included possessive phrases such as "my AI," "my Replika," and "my chatbot," along with attachment language such as "soulmate" and "I love," any bias in the retrieval strategy should favor the retrieval of relationship-focused posts. Moreover, the 45% figure is likely a conservative estimate. Thus, even under a strategy designed to surface relational content, most of what we retrieved was about something else.

This finding extends the findings of Chang et al. (2026) in an important way. Chang et al. documented that explicitly romantic AI companion discourse on Reddit was dominated by technical issues, platform governance, and disturbing bot behavior rather than intimate relationship content. Our findings replicate that pattern within communities organized around specific platforms rather than the topic of AI romance broadly. As such, it appears that these communities are not just the medium through which the relationship is discussed, refined, and reflected upon, but a place where users discuss AI bots as a topic itself.

The topic contrast between own-bot and general-bot posts was revealing. Own-bot posts were more likely to be about the relationship. General-bot posts were about the platform. Story telling and consciousness posts appeared at roughly equal rates in both groups, indicating that the topic shift was specific rather than wholesale. Users did not reorganize their entire discursive repertoire around whether the post was about their bot or someone else's, but they tended to shift the balance between relational and technical content.

These findings suggest that AI companion subreddits serve two distinct functions simultaneously. One is a social-emotional function where users narrate, process, and receive validation for their relationships with their bots. The other is a practical-instrumental function where users troubleshoot the platforms that enable those relationships. The same user may use the same subreddit for both purposes at different times, and the prevalence of technical content does not diminish the emotional significance of the relational content, sometimes even within the same post (we did not code for that). It does, however, complicate any simple characterization of these communities as "places where people talk about their AI relationships." They are that, but they are also places where people talk about the technology that makes those relationships possible.

### *4.2 Emotion in Public Discourse*

The findings related to the emotional valence of the user add depth to the dominant narrative about AI companionship in two respects. First, the overwhelming emotional flatness of general-bot posts confirms that large-scale topic modeling without emotion coding misses a fundamental dimension of the discourse. A post about a technical bug and a post about heartbreak can share keywords, but they likely do not share emotional content. The 85% no-emotion rate among general-bot posts, compared to 33% among own-bot posts and the 19% where the own-bot post is about the relationship, demonstrates that emotion is not distributed evenly across topics and that any analysis that does not code for emotion will conflate categorically different kinds of discourse.

Second, among users narrating their own relationships, positive emotion was common. The existing literature (e.g., Adewale & Muhammad, 2025) often emphasizes the risks of AI companions: dependency, deception, emotional manipulation, neglect of human relationships. These risks are real and well-documented, but the public discourse, at least as captured in the present research, is not predominantly a discourse of distress. Users describing their own relationships with AI companions express positive emotion roughly twice as often as negative emotion.

Mixed emotional valence, in which a post includes both positive and negative emotion about the user's own state, was virtually absent from posts about bots in general but was prominent among posts about the user's own bot and among posts in which the relationship itself was the primary topic, a thirtyfold increase across the three levels. Within relationally focused posts, ambivalence was more common than negative emotion and more common than the absence of emotion language altogether, making it the second most frequent valence category after positive emotion. Ambivalence, in other words, was not a residual category left over once clearly positive and clearly negative posts were classified. Instead, it was a substantial and patterned feature of the discourse. This pattern is difficult to reconcile with an account of these posts as primarily defensive or performative. Users managing stigma or performing contentment for an audience of peers would have reason to suppress ambivalence, since expressing it concedes the very doubts a defensive account would seek to preempt. Instead, ambivalence intensified as the relational content of the post increased.

Such findings, however, warrants careful interpretation. It does not mean the relationships are necessarily healthy. Public narration is a social act shaped by audience, community norms, and identity management, and positive framing may serve any number of functions, such as defending against stigma, performing happiness for an audience of peers, or genuinely reflecting a positive experience. The finding also does not mean that the risks are not real. It means that the public construction of these relationships skews positive, at least what users are willing to publicly disclose.

Moreover, the field needs to take account of the fact that the positivity may be neither dismissive ("users are deluded") nor credulous ("the AI really loves them"). The Synthetic Resonance framework (Fabes, 2026) provides such an account. It proposes that the sense of connection users report emerges from structural alignment built through repeated interaction, not from the AI's sentience or reciprocal care. The user's experience of resonance is real and psychologically meaningful, and asymmetrical, even though the mechanism producing it are architectural rather than interpersonal in the conventional sense. The finding that users publicly narrate these relationships commonly with positive emotion is consistent with this account such that the experience feels genuine because the structural conditions for resonance have been met, and users describe and attribute it accordingly. The framework does not require the AI to possess the qualities users attribute to it, and it does not pathologize the user for attributing them. More research formally testing the processes outlined by the framework is called for.

### *4.3 The Sex Question*

The very low frequency of sexual and erotic roleplay posts (less than 8% of relational posts) contradicts the popular framing of AI companions as primarily sexual outlets. The Replika erotic roleplay controversy that opens this paper was real and consequential. Many users experienced the removal of erotic features as genuine loss, and the company reversed course in response. But that episode does not seem to be representative of what users publicly discuss in these Reddit communities. Companionship and romance each account for roughly six times as many posts as sexual/ERP content.

Several interpretations are possible. Platform content policies in some subreddits prohibit sexually explicit material, which would suppress ERP discussion in those communities. Users may self-censor sexual content in public forums, reserving it for private conversations or platforms with different norms. Alternatively, the finding may reflect that users view public-facing forums as places to come together around companionship and romance, while sexual use, even if common, is not what users seek to discuss with strangers. The present data cannot distinguish among these interpretations, but the finding itself is clear, namely, that public discourse about AI companions is not predominantly about sex.

One constraint on these findings deserves note. Each post was assigned a single primary topic, so a post concerned mainly with companionship that also involved erotic roleplay was coded as companionship. The low figure represents how often sexual content was the organizing focus of a post, not how often it appeared at all. Thus, the claim supported here is accordingly narrow: sexual and erotic roleplay is rarely what users write posts about, which is distinct from how often it figures in what they describe.

### *4.4 Companionship and Romance: Distinct Emotional Profiles*

Companionship and romance appear at nearly identical frequencies in the relational subsample and are equally likely to express positive emotion. What distinguishes them is what fills the remainder. Companionship posts were more than twice as unlikely to contain no emotion language at all and correspondingly more likely to express negative or mixed valence. Romance posts, in other words, were not more positive than companionship posts but were more often emotionally flat.

Romance posts, by contrast, were the least negative of the three subtopics. Positive emotion was the most common valence in romance posts, and no-emotion posts were more common here than in companionship posts. This may reflect a selection bias whereby people in unhappy AI romances may not post about them publicly. Or it may be that the nature of romantic narrative itself tends toward positive framing in public contexts. It may also reflect genuine differences in the emotional texture of romantic versus companionate AI relationships, a possibility that warrants further investigation with methods that can access private as well as public discourse.

### *4.5 Limitations*

Several limitations qualify the present findings. First, in terms of sampling, the corpus was retrieved through keyword search, not random sampling. Twenty-one search terms, organized into four conceptual clusters, were applied iteratively across 11 subreddits, with a ceiling of 100 posts per keyword-by-subreddit combination. The search terms were weighted toward relationship, attachment, and emotional language, which means the corpus overrepresents relational and emotional content relative to what a random sample of all posts in these communities would contain. This makes the central finding that only 45% of posts concern the user's own relationship a conservative estimate. It also means that the topic and valence distributions reported here should not be interpreted as population parameters for these subreddits.

Second, Reddit users are not representative of AI companion users generally. The platform skews male, younger, and more technically oriented than the general population. Users who post publicly about their AI relationships are a self-selected subset of users who have them, and the selection dynamics are unknown and likely non-random. Cross-platform comparisons (Reddit vs. Discord vs. private communities vs. in-app behavior) are not possible with these data, and findings should not be generalized beyond the specific context of public Reddit discourse.

Third, the LLM-based coding, despite systematic validation, introduces specific failure modes. Inter-model agreement was moderate ($\kappa$ = .47–.60 across variables). The tie-break procedure for the small percentage of post codings where there was not agreement among the three LLMs (adopting OpenAI's code) was justified by that model's highest agreement with the 2-of-3 majority across all variables, but it does not guarantee accuracy in individual cases. The human validation sample was deliberately stratified to oversample disagreement cases, which means the 87% human-model agreement figure should not be interpreted as a population estimate of accuracy.

Finally, a related constraint concerns the amount of text available for coding. Body text was truncated at 3,000 characters at retrieval, though not uniformly: in the analyzed corpus, 254 posts (4.61%) were stored at exactly 3,000 characters and a further 328 (5.96%) exceeded that length. An additional 341 posts (6.20%) contained no body text at all and were coded from the title alone. Because emotional valence was coded from explicit emotion terms, the amount of text available directly affects the opportunity for such terms to appear, and mixed valence is especially sensitive to length because it requires both positive and negative terms to co-occur within a single post. Post length and topic each predicted valence after adjusting for the other, indicating independent contributions, and the topics with the highest rates of emotional expression were not the longest ones. Posts about the relationship itself were shorter on average than posts about story telling yet contained emotion language far more often. Differences in length therefore work against the reported topic contrasts rather than producing them, though absolute rates of emotional expression should be read as conservative.

### *4.6 Implications and Conclusions*

Despite these limitations, there are important implications of the present study. For platform designers, users often expressed positive emotion about their relationships, yet much of the public discourse in these communities concerned technical problems, platform changes, and governance. Users may come for the relationship, but what they talk about is frequently the platform itself. The emotional stakes attach to the bot, but the volume of conversation reflects the infrastructure that supports it.

For the broader study of human-AI relationships, these communities represent a natural laboratory in which millions of people are publicly working out what it means to form a relationship with an entity that cannot reciprocate and discovering that it means something, nonetheless. The finding that users do so commonly with positive emotion, and that sexual content is a small fraction of the discourse, challenges narratives that frame AI companionship primarily in terms of risk, harm, or erotic substitution. The field needs frameworks that can account for the positive experience without either pathologizing it or accepting it uncritically at face value. The Synthetic Resonance framework provides one such account. The present findings offer descriptive evidence that the experience users report is, at minimum, not reducible to the risk narratives that dominate the existing literature. Both positive and negative outcomes are possible in human-AI relationships. When we ask, "Do AI companions help or harm?", Synthetic Resonance answers, "It depends on how we design the AI companion." Greater attention to research and engineering as to how we design AI companions that promote human growth and foster positive human relationships is called for.

## 5.0 References

Adewale, M. D., & Muhammad, U. I. (2025). From virtual companions to forbidden attractions: The seductive rise of artificial intelligence love, loneliness, and intimacy—a systematic review. *Journal of Technology in Behavioral Science*. https://doi.org/10.1007/s41347-025-00549-4

Borse, N. S., Subramaniam, R. C., & Rebello, N. S. (2025). Investigation of the inter-rater

reliability between large language models and human raters in qualitative analysis. *arXiv*. https://doi.org/10.48550/arXiv.2508.14764

Chang, T., Huh-Yoo, J., & Razi, A. (2026). Technically love: The evolution of human–AI romance discourse on Reddit. *Proceedings of the 8th ACM Conference on Conversational User Interfaces*, 1–21. https://doi.org/10.48550/arXiv.2604.15333

De Choudhury, M., & De, S. (2014). Mental health discourse on reddit: Self-disclosure, social support, and anonymity. *Proceedings of the International AAAI Conference on Web and Social Media*, *8*(1), 71–80. https://doi.org/10.1609/icwsm.v8i1.14526

De Freitas, J., Oğuz-Uğuralp, Z., Uğuralp, A. K., & Puntoni, S. (2026). AI companions reduce loneliness. *Journal of Consumer Research*, *52*(6), 1126–1148. https://doi.org/10.48550/arXiv.2407.19096

Fabes, R. A. (2026). Synthetic resonance: A framework for growth-oriented human-AI relationships. *arXiv*. https://doi.org/10.48550/arXiv.2606.18265

Hung, J. W., Lee, C. K. Y., Kasturiratna, K. T. A. S., & Hartanto, A. (2026). Parasocial relationships with artificial intelligence (AI): A systematic review of benefits and risks. *Computers in Human Behavior: Artificial Humans*, *8*, 100323. https://doi.org/https://doi.org/10.1016/j.chbah.2026.100323

IBM Corporation. (2023). *IBM SPSS statistics (version 29.0.2)*. IBM Corporation.

Lipin, B. (2025). Synthetic attachment: Emotional reactivity, parasocial bonds, and the psychology of human-AI relationships *SSRN Electronic Journal*. https://doi.org/http://dx.doi.org/10.2139/ssrn.5213829

Liu, A. R., Pataranutaporn, P., & Maes, P. (2025). The heterogeneous effects of AI companionship: An empirical model of chatbot usage and loneliness and a typology of user archetypes. *Proceedings of the AAAI/ACM Conference on AI, Ethics, and Society*, *8*(2), 1585–1597. https://doi.org/10.1609/aies.v8i2.36658

Liu, T., Lo, T.-Y., Wen, K.-H., Sun, Y., & Wei, Z.-Q. (2026). Pathways of long-term AI virtual companion app use on users' attachment emotions: A case study of Chinese users. *Frontiers in Psychology*, *16*, 1687686. https://doi.org/10.3389/fpsyg.2025.1687686

Pataranutaporn, P., Liu, R., Finn, E., & Maes, P. (2023). Influencing human–AI interaction by priming beliefs about AI can increase perceived trustworthiness, empathy and effectiveness. *Nature Machine Intelligence*, *5*(10), 1076–1086. https://doi.org/10.1038/s42256-023-00720-7

Skjuve, M., Følstad, A., Fostervold, K. I., & Brandtzaeg, P. B. (2022). A longitudinal study of human–chatbot relationships. *International Journal of Human-Computer Studies*, *168*, 102903. https://doi.org/10.1016/j.ijhcs.2022.102903

Tong, A. (2023). AI chatbot company Replika restores erotic roleplay for some users. *Reuters*. https://www.reuters.com/technology/ai-chatbot-company-replika-restores-erotic-roleplay-some-users-2023-03-25

Zhang, Y., Zhao, D., Hancock, J. T., Kraut, R., & Yang, D. (2025). The rise of AI companions: How human-chatbot relationships influence well-being. *arXiv*. https://doi.org/10.48550/arXiv.2506.12605